\documentclass{andromedaone}       
\usepackage{amssymb}
\usepackage{amsmath}
\usepackage{float}
\usepackage{hyperref}

\restylefloat{table}

\journal{BSM}
\vol{2021}
\jyear{Egypt}
\pages{Zewail City of Science, Technology and Innovation} 

\received{xx January 2018}
\published{xx March 2018}

\def\be{\begin{equation}}
\def\ee{\end{equation}}
\def\bea{\begin{eqnarray}}
\def\eea{\end{eqnarray}}

\begin{document}

\title{Trimaximal mixing with one texture zero from type II seesaw and $\Delta(54)$  family symmetry}

\author{M. A. Loualidi\footnote{mr.medamin@gmail.com}}
\address{LPHE-Modeling and Simulations, Faculty of Science, Mohammed V University in Rabat, 10090, Morocco}

\begin{abstract}
We propose a neutrino model based on a $\Delta(54)$ flavor symmetry suitable for explaining the current neutrino oscillation data. Neutrino masses arise from the type II seesaw mechanism where the $\Delta(54)$ field assignments has led to a simple neutrino mass matrix with one texture zero and which satisfies the magic symmetry consistent with the well-known
trimaximal mixing matrix. We found interesting predictions concerning neutrino masses and mixing. In particular, only the normal neutrino mass hierarchy and the lower octant of the atmospheric angles are allowed in this model. The model predicts as well that the $CP$ conserving values for the Dirac $CP$ phase $\delta _{CP}$ are not allowed and thus, $CP$ is always violated in the neutrino sector. We have also investigated the sum of absolute neutrino masses from cosmological observations, the effective Majorana mass $m_{\beta \beta }$\ from neutrinoless double beta decay experiments, and the electron neutrino mass $m_{\beta }$\ from beta decays where we found that the obtained range of $m_{\beta \beta }$\ can be tested by several experiments in the near future.
\end{abstract}

\maketitle

\begin{keyword}
Neutrino model building\sep Trimaximal mixing\sep One texture zero\sep Flavor symmetries.
\end{keyword}

\section{Introduction}
Despite the major progress in neutrino physics over the last two
decades, there are many questions and properties related to
neutrinos that are yet to be addressed such as:
$\mathbf{(i)}${\normalsize \ }whether neutrinos are their own
antiparticles or not, which is equivalent to asking whether these
particles are Majorana or Dirac fermions; $\mathbf{(ii)}$ the
presence of $CP$ violation in the chargeless sector;
$\mathbf{(iii)}$ the absolute scale of neutrino masses;{\normalsize
	\ }$\mathbf{(iv)}${\normalsize \ }the neutrino mass ordering
problem; and $\mathbf{(v)}$ the octant of the atmospheric mixing
angle.{\normalsize \ }The investigation of these features provides
us with a clear evidence that we need to look beyond the standard
model (SM) of electroweak interactions.{\normalsize \newline }There
are dozens of experiments looking at neutrinos from different angles
\cite{R1,A1}, each using different insights to help unraveling the
properties of this elusive particle. Theoretically, one of the best
ways to generate neutrino masses is via the so-called seesaw
mechanism where the smallness of neutrino mass is explained by the
presence of super-heavy partners. There are three popular
realizations of this mechanism: type I
seesaw with right-handed neutrinos \cite{B1}, type II seesaw with heavy $%
SU(2)_{L}$ scalar triplets \cite{R2} and type III see-saw with $SU(2)_{L}$
fermion triplets \cite{B2}\footnote{%
	One must note that small neutrino masses can also be generated through the
	so-called scotogenic mechanism \cite{B3} whereby the neutrino mass problem
	is connected to dark matter. On the other hand, the phenomenology of
	scotogenic models is interesting and the new extra degrees of freedom can be
	tested at present and future TeV-scale colliders \cite{B4}. Besides, the
	problem of neutrino mixing can be addressed as usual by the introduction of
	discrete symmetry groups, see for instance Ref. \cite{B5} and the references
	therein.}. On the other hand, one of the most popular patterns of the
leptonic mixing is acquired by the trimaximal (TM$_{2}$) mixing matrix which
predicts non-zero reactor angle $\theta _{13}$ and non-maximal atmospheric
angle $\theta _{23}$. This mixing matrix is defined as \cite{R3}%
\begin{equation}
	U_{TM_{2}}=\left(
	\begin{array}{ccc}
		\sqrt{\frac{2}{3}}\cos \theta & \frac{1}{\sqrt{3}} & \sqrt{\frac{2}{3}}\sin
		\theta e^{-i\sigma } \\
		-\frac{\cos \theta }{\sqrt{6}}-\frac{\sin \theta }{\sqrt{2}}e^{i\sigma } &
		\frac{1}{\sqrt{3}} & \frac{\cos \theta }{\sqrt{2}}-\frac{\sin \theta }{\sqrt{%
				6}}e^{-i\sigma } \\
		-\frac{\cos \theta }{\sqrt{6}}+\frac{\sin \theta }{\sqrt{2}}e^{i\sigma } &
		\frac{1}{\sqrt{3}} & -\frac{\cos \theta }{\sqrt{2}}-\frac{\sin \theta }{%
			\sqrt{6}}e^{-i\sigma }%
	\end{array}%
	\right) .U_{P}  \label{tm2}
\end{equation}%
where $\theta $ parameterizes the deviation from the well known tribimaximal
(TBM) ansatz \cite{R4}, $\sigma $ is an arbitrary phase and $U_{P}=\mathrm{%
	diag}\left( 1,e^{i\alpha _{21}/2},e^{i\alpha _{31}/2}\right) $ is a diagonal
matrix that encodes the Majorana phases $\alpha _{21}$ and $\alpha _{31}$.
In this regard, it is commonly known that the observed pattern of neutrino
masses and mixing suggest the existence of symmetries acting in the space of
fermion generations. In particular, the TBM mixing can be derived using
non-Abelian discrete symmetries and subsequently,\textrm{\ }TM$_{2}$ is
obtained by rotating the TBM matrix in the 1-3 plane. Moreover, one of the
simplest ways to reduce the number of free parameters is by developing
neutrino mass matrices with one texture zero \cite{A2,A3,A4}. Realizing
these types of matrices leads to special correlations among the physical
quantities such as mass squared differences, mixing angles\ and $CP$ phases
and hence\textrm{\ }allows for highly predictive models; see Ref. \cite{A5}
for models studying neutrino mass matrices with one texture zero and TM$_{2}$%
.

Here we present a model where the neutrinos acquire their masses through the
type II seesaw mechanism whereas the neutrino flavor structure is provided
by the $\Delta (54)$\ flavor group. In this setup, three gauge singlet
scalar fields (flavons) are required by $\Delta (54)$ invariance and
phenomenological reasons. Furthermore, the leptonic mixing arises from the
neutrino sector and thus, the $\Delta (54)$ charge assignments of lepton and
scalar fields are chosen in a way that the resulting charged lepton Yukawa
matrix is diagonal. Afterwards, the $\Delta (54)$ flavor symmetric
construction with a specific choice of flavon alignment leads to highly
predictive results and give rise to a neutrino mass matrix with two
important features: \emph{(a)} one texture zero reducing the number of free
parameters to only three; and \emph{(b) }satisfies the magic symmetry and
thus diagonalized by TM$_{2}$ mixing matrix in (\ref{tm2}), see e.g., Ref.
\cite{R5}. In particular, we concluded that only the normal neutrino mass
hierarchy is permitted in our model, while we showed through scatter plots
that the atmospheric angle lies in the lower octant. Moreover, it is found
that the $CP$ conserving values for the Dirac $CP$ phase $\delta _{CP}$ are
not allowed in our model and thus, $CP$ is always violated in the neutrino
sector. On the other hand, the phenomenological implications of the light
neutrino masses are explored using recent global fit of neutrino oscillation
parameters from Ref. \cite{R6}.\textrm{\ }In particular, we have
investigated the sum of the three active neutrino masses $\sum m_{i}$\ from
cosmological observations, the effective Majorana mass $\left\vert m_{\beta
	\beta }\right\vert $\ from neutrinoless double beta decay ($0\nu \beta \beta
$) experiments, and the electron neutrino mass $m_{\beta }$\ from beta
decays. It is found that the obtained ranges of $\sum m_{i}$\ and $%
\left\vert m_{\beta \beta }\right\vert $\ can be tested by different planned
future experiments while the predicted range of $m_{\beta }$\ requires more
enhanced sensitivity of beta decay experiments.
This contribution is structured as follows. In section 2, we give some
properties of the $\Delta (54)$ group, then we present our $\Delta (54)$
field assignments. In section 3, we examine in details the neutrino
oscillation parameters in our model. In section 3, we study the
phenomenological consequences of the neutrino masses from non-oscillatory
experiments. In section 4, we give our conclusion.
\section{Implementation of $\Delta (54)$ flavor symmetry}

In this section, we describe the particle content and the introduction of $%
\Delta (54)$ flavor symmetry in the SM. Before giving the full field
spectrum of our model, let us first mention some properties of the $\Delta
(54)$ group. First, notice that the discrete flavor $\Delta (54)$\ symmetry
has ten irreducible representations $\mathbf{R}_{i}$\ where in addition to
two singlets $\mathbf{1}_{+}$ and $\mathbf{1}_{-}$, and four doublets $%
\mathbf{2}_{1,2,3,4}$, it has four three-dimensional irreducible
representations denoted as $\mathbf{3}_{k}$ and their conjugates $\overline{%
	\mathbf{3}}_{k}$ with index $k=1,2$. The tensor products between the
representations that are\ relevant to the present work are as follows%
\begin{eqnarray}
	\mathbf{1}_{+}\otimes \mathbf{R}_{i} &=&\mathbf{R}_{i}\quad ,\quad \mathbf{1}%
	_{-}\otimes \mathbf{1}_{-}=\mathbf{1}_{+}\quad ,\quad \mathbf{1}_{-}\otimes
	\mathbf{3}_{1}=\mathbf{3}_{2}\quad ,\quad \mathbf{1}_{-}\otimes \mathbf{\bar{%
			3}}_{1}=\mathbf{\bar{3}}_{2}  \nonumber \\
	\mathbf{1}_{-}\otimes \mathbf{3}_{2} &=&\mathbf{3}_{1}\quad ,\quad \mathbf{1}%
	_{-}\otimes \mathbf{\bar{3}}_{2}=\mathbf{\bar{3}}_{1}\quad ,\quad \mathbf{3}%
	_{2}\otimes \mathbf{\bar{3}}_{2}=\mathbf{1}_{+}\oplus \mathbf{2}_{1}\oplus
	\mathbf{2}_{2}\oplus \mathbf{2}_{3}\oplus \mathbf{2}_{4}  \nonumber \\
	\mathbf{3}_{1}\otimes \mathbf{\bar{3}}_{1} &=&\mathbf{1}_{+}\oplus \mathbf{2}%
	_{1}\oplus \mathbf{2}_{2}\oplus \mathbf{2}_{3}\oplus \mathbf{2}_{4}\quad
	,\quad \mathbf{\bar{3}}_{2}\otimes \mathbf{3}_{1}=\mathbf{1}_{-}\oplus
	\mathbf{2}_{1}\oplus \mathbf{2}_{2}\oplus \mathbf{2}_{3}\oplus \mathbf{2}_{4}
	\nonumber \\
	\mathbf{\bar{3}}_{1}\otimes \mathbf{3}_{2} &=&\mathbf{1}_{-}\oplus \mathbf{2}%
	_{1}\oplus \mathbf{2}_{2}\oplus \mathbf{2}_{3}\oplus \mathbf{2}_{4}\quad
	,\quad \mathbf{3}_{2}\otimes \mathbf{3}_{2}=\mathbf{\bar{3}}_{2}\oplus
	\mathbf{\bar{3}}_{2}\oplus \mathbf{\bar{3}}_{1}  \nonumber \\
	\mathbf{3}_{1}\otimes \mathbf{3}_{1} &=&\mathbf{\bar{3}}_{2}\oplus \mathbf{%
		\bar{3}}_{2}\oplus \mathbf{\bar{3}}_{1}  \label{tp}
\end{eqnarray}%
A complete list of tensor products of $\Delta (54)$ irreducible
representations can be found in \cite{R7}. Moreover, this non-Abelian group
is isomorphic to the semidirect product $\mathbb{S}_{3}\times \mathbb{Z}%
_{3}\times \mathbb{Z}_{3}$, where $\mathbb{S}_{3}$\ is the symmetric group
of degree three. Therefore, $\Delta (54)$ is generated by four elements; two
generators $\mathcal{S}$ and $\mathcal{T}$ for the $\mathbb{S}_{3}$ group
and the remaining two, denoted as $\mathcal{U}$ and $\mathcal{V}$, generate $%
\mathbb{Z}_{3}\times \mathbb{Z}_{3}$. These four generators satisfy the
following identity%
\begin{equation}
	\mathcal{S}^{3}=\mathcal{T}^{2}=\left( \mathcal{ST}\right) ^{2}=\mathcal{U}%
	^{3}=\mathcal{V}^{3}=\mathbf{{I}_{id}}
\end{equation}%
This group can be utilized as a flavor symmetry, but is rarely invoked in
particle physics model building; see reference \cite{A6} for\textrm{\ }a
simple extension of the SM using $\Delta (54)$\ providing a good description
of the neutrino sector.\newline
Now, we present our field content\ where in our extension with a $\Delta (54)
$ symmetry, it can be arranged into two sets: $\emph{(i)}$ the usual lepton
and Higgs fields of the SM but carrying as well charges under $\Delta (54)$;
these are the three lepton doublets $L_{i}$\ with $L_{1}=(\nu
_{e},e^{-})_{L},$ $L_{2}=(\nu _{\mu },\mu ^{-})_{L}$ and $L_{3}=(\nu _{\tau
},\tau ^{-})_{L}$ which are assigned to the irreducible triplet $\mathbf{3}%
_{2}$ of $\Delta (54)$, the three right-handed leptons $E_{i}=(e_{R},\mu
_{R},\tau _{R})$ assigned to the $\Delta (54)$ triplet $\mathbf{\bar{3}}_{2}$%
, and finally the Higgs doublet $H=(H^{+},H^{0})$ which transforms as a
trivial singlet under the $\Delta (54)$ group. \emph{(ii)} An extra scalar
sector that contains an $SU(2)_{L}$ scalar triplet and flavon fields
transforming as gauge singlet which are required by $\Delta (54)$ flavor
invariance and phenomenological purposes. The $SU(2)_{L}$ triplet $T$\ is
responsible for the generation of neutrino masses via type II seesaw
mechanism while we have added in total three flavons denoted as $\chi $, $%
\phi $\ and $\rho $;\textrm{\ }the first two\ are needed to produce a mass
matrix compatible with the the familiar TBM matrix while the third flavon $%
\rho $ is needed to produce a neutrino mass matrix with magic symmetry; that
is a matrix which is compatible with neutrino mixing matrix of trimaximal
form. The $\Delta (54)$ quantum numbers that are relevant for this work are
summarized in table (\ref{t1}).
\begin{table}[h!]
	\centering
\begin{tabular}{c|r|r||r|r|r|r|r}
\hline\hline
\textbf{Fields} & $L_{i}=(L_{1},L_{2},L_{3})$ & $E_{i}=\left( e_{R},\mu
_{R},\tau _{R}\right) $ & $H$ & $T$ & $\chi $ & $\phi $ & $\rho $ \\
 \hline
$\Delta (54)$ & $\mathbf{3}_{2}$ & $\mathbf{\bar{3}}_{2}$ & $\mathbf{1}_{+}$
& $\mathbf{1}_{-}$ & $\mathbf{3}_{2}$ & $\mathbf{3}_{1}$ & $\mathbf{1}_{-}$
\\ \hline\hline
\end{tabular}
\caption{Quantum numbers of lepton and scalar fields under $\Delta(54)$.}
\label{t1}
\end{table}
As mentioned in the introduction, we will mainly concentrate on the neutrino
sector; thus, we perform the present study in the basis where the charged
lepton mass matrix is diagonal. In this scenario, the leptonic mixing matrix
is the one that diagonalizes the light neutrino mass matrix.\textrm{\ }%
Indeed, with the field assignments given in table (\ref{t1}) the Lagrangian
of the charged leptons given by
\begin{equation}
	\mathcal{L}_{Y}^{l}=\sum_{l=e,\mu ,\tau }Y_{l}^{ij}\overline{L}%
	_{i}E_{j}^{c}H+h.c.  \label{E}
\end{equation}%
leads, after performing the relevant $\Delta (54)$ tensor product\footnote{%
	This tensor product contains the $\Delta (54)$ trivial singlet $\mathbf{1}_{%
		\mathbf{+}}$ as can be checked from eq. (\ref{tp}).} $\mathbf{\bar{3}}%
_{\left( 0,-1,0\right) }\otimes \mathbf{3}_{\left( 0,-1,0\right) }\otimes
\mathbf{1}_{\left( 1,1,1\right) }$ and electroweak symmetry breaking, to a
diagonal mass matrix $\mathcal{M}^{l}=\mathrm{diag}(m_{e},m_{\mu },m_{\tau
})=\frac{\upsilon _{H}}{\sqrt{2}}\mathrm{diag}(Y_{e},Y_{\mu },Y_{\tau })$
with $\upsilon _{H}$ being the vacuum expectation value (VEV) of the Higgs
doublet $H$.\ The hierarchy among these masses can be obtained by invoking
the well-known Froggatt-Nielsen mechanism, see Ref. \cite{R8} for more
details.

\section{Neutrino masses and mixing}

\begin{figure}
	\centering
	\hspace{3em}\includegraphics[width=.44\textwidth]{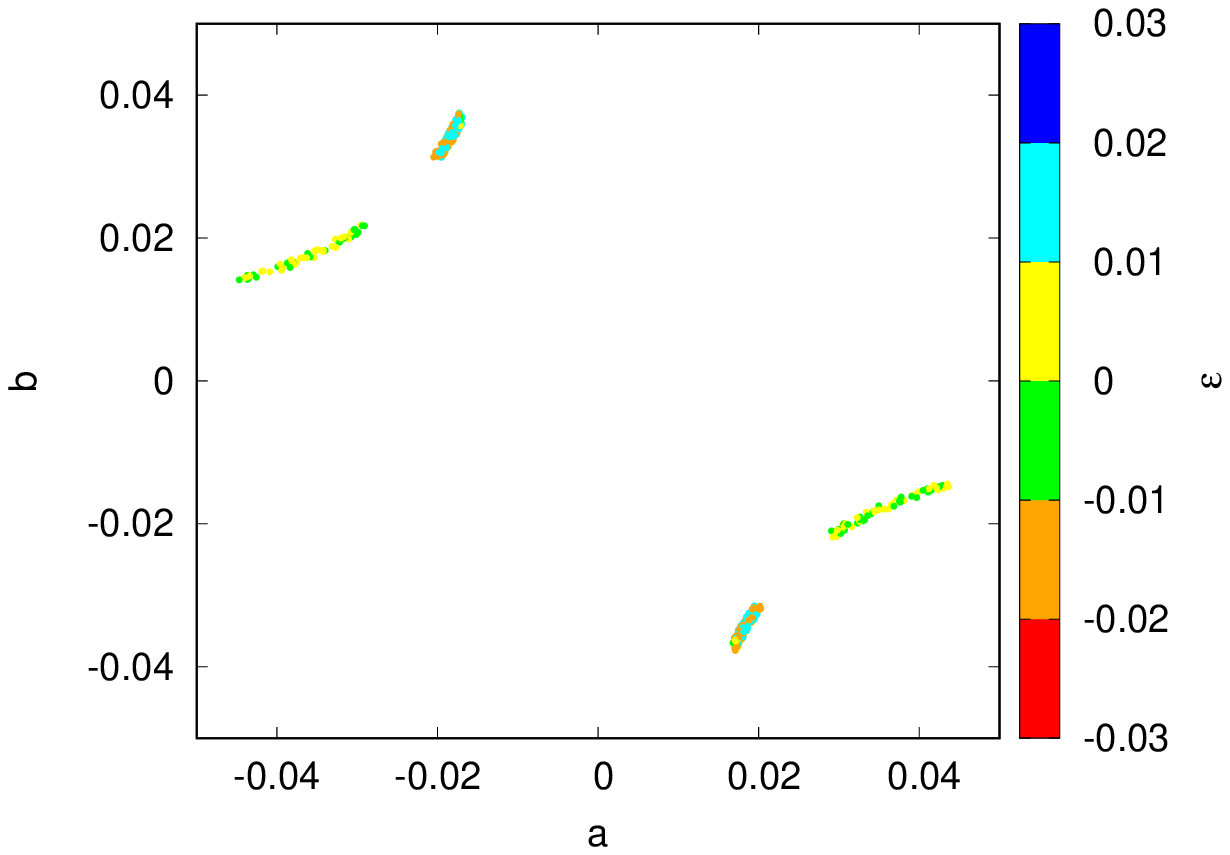}\quad %
	\includegraphics[width=.44\textwidth]{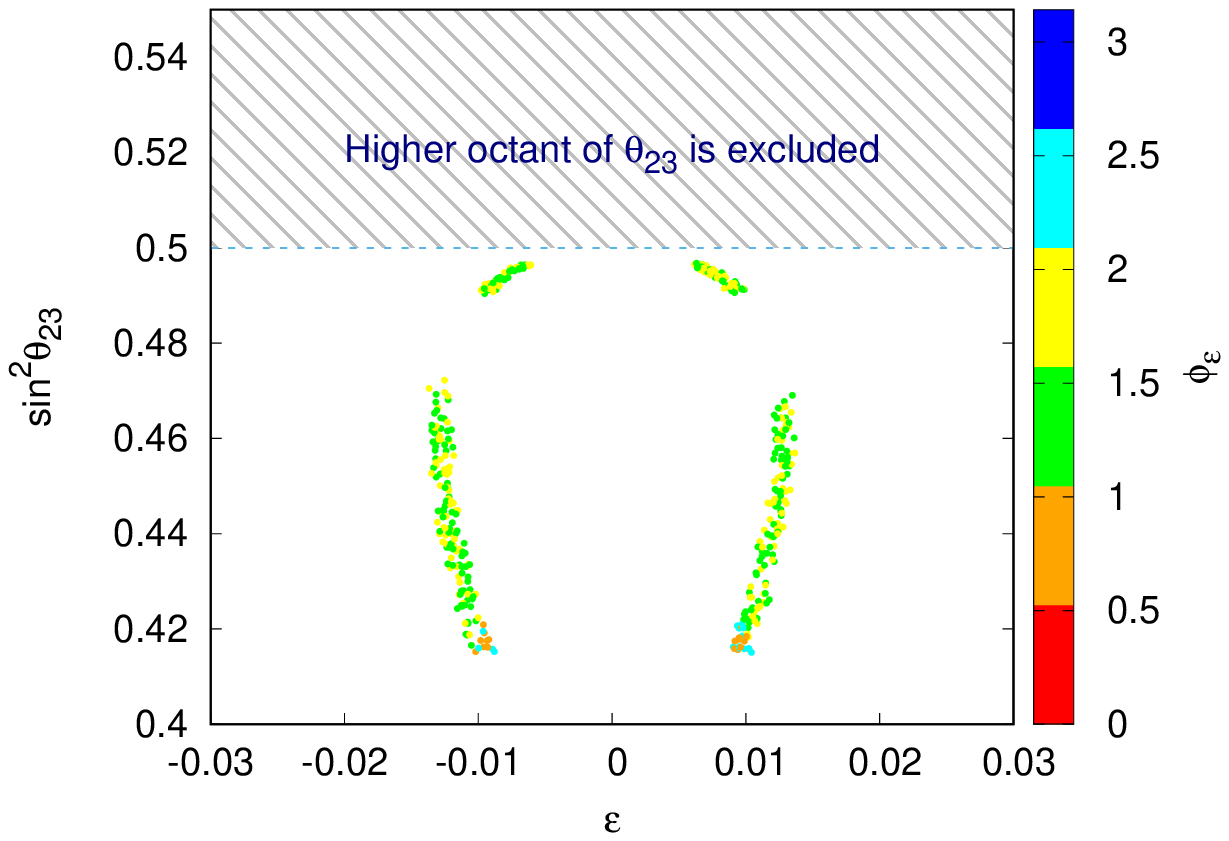}
	\par
	\caption{Left: The correlation among the parameters $a$, $b$ and $\protect%
		\varepsilon $. Right: Scatter plot on the plan of $\sin ^{2}\protect\theta %
		_{23}$\ and $\protect\varepsilon $ with the palette showing $\protect\phi _{%
			\protect\varepsilon }$.}
	\label{01}
\end{figure}

\begin{figure}
	\centering
	\hspace{2.5em} \includegraphics[width=.44\textwidth]{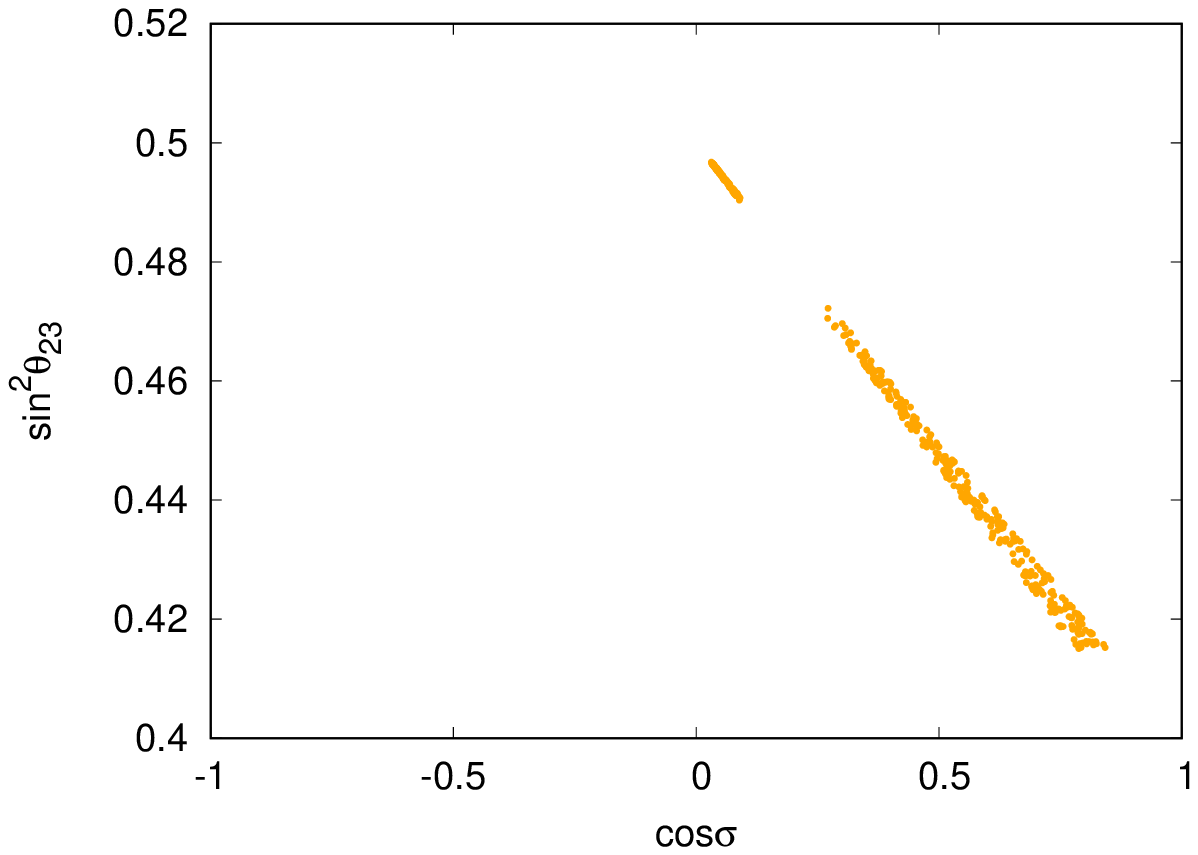}\quad %
	\includegraphics[width=.44\textwidth]{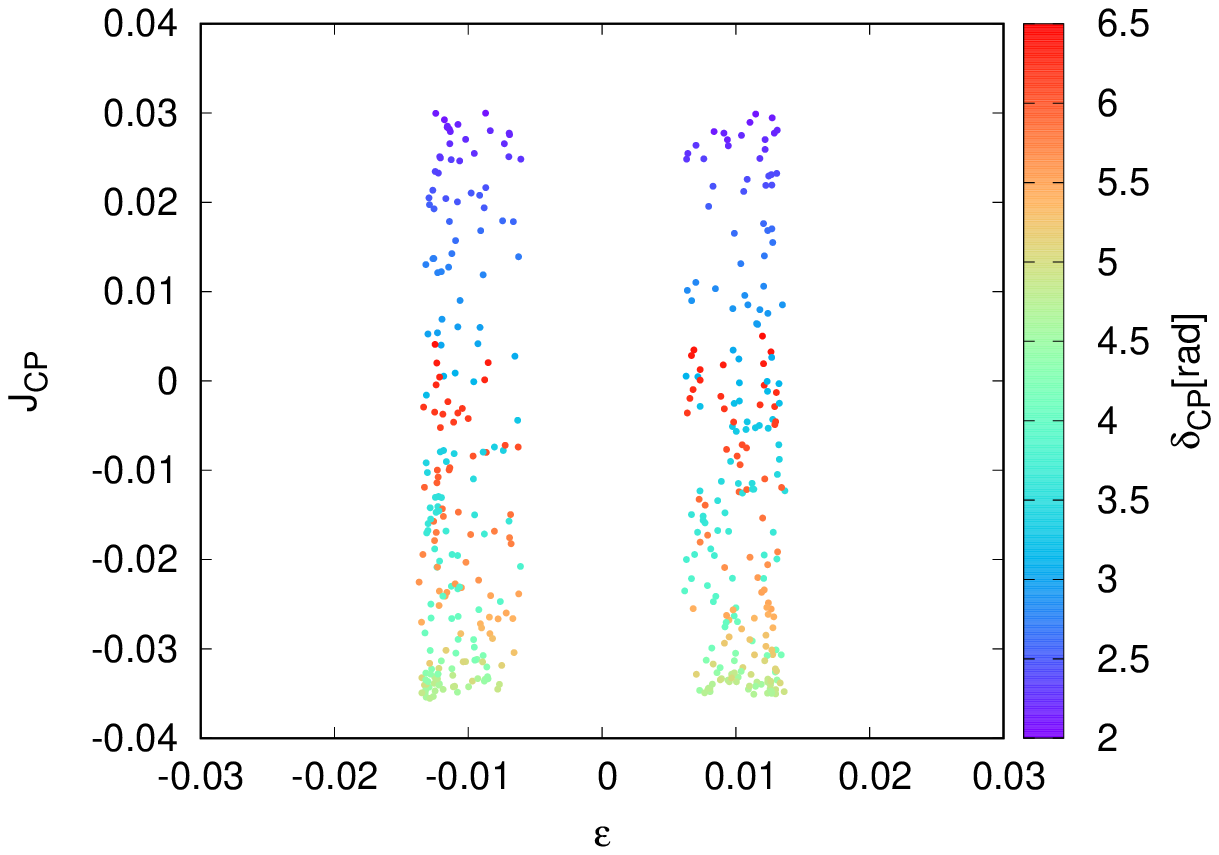}
	\par
	\caption{Left: The variation of $\sin \protect\theta _{23}^{2}$ as a
		function of \ $\cos \protect\sigma $. Right: Scatter plot on the plan of $%
		J_{CP}$\ and $\protect\varepsilon $ with the palette showing the Dirac $CP$
		phase $\protect\delta _{CP}$.}
	\label{02}
\end{figure}

In this section, we build a predictive neutrino model with one texture zero
mass matrix based on the $\Delta (54)$ non-Abelian flavor symmetry. The
light neutrino masses are generated via the type II seesaw mechanism which
requires extending the scalar sector of the SM by a scalar triplet $\vec{T}$
with hypercharge $Y\vec{T}=2\vec{T}$, and which may be represented by the
following traceless complex $2\times 2$ matrix%
\begin{equation}
	T=\vec{\sigma}\vec{T}=\left(
	\begin{array}{cc}
		\frac{T^{+}}{\sqrt{2}} & T^{++} \\
		T^{0} & -\frac{T^{+}}{\sqrt{2}}%
	\end{array}%
	\right)  \label{2.1}
\end{equation}%
where $\sigma ^{a}$\ are $2\times 2$ Pauli matrices. Moreover, we add three
flavon fields---$\chi $, $\phi $ and $\rho $---carrying quantum numbers
under the $\Delta (54)$ symmetry. Now, by using the particle assignments
shown in table \ref{t1}, the $\Delta (54)$-invariant Lagrangian for
neutrinos is given by
\begin{equation}
	-\mathcal{L}_{\nu }^{II}=\frac{Y_{\chi }^{ij}}{\Lambda }\left( \overline{L}%
	_{i}^{c}i\sigma _{2}TL_{j}\right) \chi +\frac{Y_{\phi }^{ij}}{\Lambda }%
	\left( \overline{L}_{i}^{c}i\sigma _{2}TL\right) _{i}\phi +\frac{Y_{\phi
			\rho }^{ij}}{\Lambda ^{2}}\left( \overline{L}_{i}^{c}i\sigma
	_{2}TL_{j}\right) \phi \rho +h.c.  \label{ss2}
\end{equation}%
After electroweak symmetry breaking and flavor symmetry breaking, the
triplet $T$ acquires a VEV $\upsilon _{T}$ by its neutral component $T^{0}$
while the flavon fields develop VEVs along the directions%
\begin{equation}
	\left\langle \chi \right\rangle =\left(
	\begin{array}{c}
		\upsilon _{\chi } \\
		0 \\
		0%
	\end{array}%
	\right) \ \ ,\ \ \left\langle \phi \right\rangle =\left(
	\begin{array}{c}
		0 \\
		\upsilon _{\phi } \\
		0%
	\end{array}%
	\right) \ \ ,\ \ \left\langle \rho \right\rangle =\upsilon _{\rho }
\end{equation}%
The resulting mass matrix for light neutrino masses is expressed as a
function of three free parameters%
\begin{equation}
	M_{\nu }^{II}=\left(
	\begin{array}{ccc}
		a+b & 0 & \varepsilon \\
		0 & \varepsilon +b & a \\
		\varepsilon & a & b%
	\end{array}%
	\right) \ \text{with\ }a=Y_{\chi }\frac{\upsilon _{T}\upsilon _{\chi }}{%
		\Lambda }\ ,\ b=Y_{\phi }\frac{\upsilon _{T}\upsilon _{\phi }}{\Lambda }\ ,\
	\varepsilon =Y_{\phi \rho }\frac{\upsilon _{T}\upsilon _{\rho }\upsilon
		_{\phi }}{\Lambda ^{2}}
\end{equation}%
This matrix has the magic symmetry as the sum of the elements in any of its
columns or rows is exactly the same. Thus, $M_{\nu }^{II}$ is diagonalized
by the well-known trimaximal mixing matrix TM$_{2}$ described by two free
parameters---an arbitrary angle $\theta $ and a phase $\sigma $ that will be
related later on to the Dirac $CP$ phase---that can be determined using the
neutrino oscillation data. To extract the mass eigenvalues, we assume that
the parameters $a\ $and $b$ are real while $\varepsilon$ is complex; $%
\varepsilon =\left\vert \varepsilon \right\vert e^{i\phi _{\varepsilon }}$.
This assumption is reasonable since in the limit where $\varepsilon%
\rightarrow 0$ the neutrino mass matrix---depending on $a$, $b$---has the
TBM form which is $CP$ conserving. Therefore, the diagonalization of $M_{\nu
}^{II}$ using the trimaximal mixing matrix defined in eq. (\ref{tm2})---$%
U_{TM}^{T}M_{\nu }^{II}U_{TM}=\mathrm{diag}(\left\vert m_{1}\right\vert
,\left\vert m_{2}\right\vert ,\left\vert m_{3}\right\vert )$---leads to the
following eigenvalues up to order\footnote{%
	See Refs. \cite{A7,A8} for more details on the diagonalization of the
	neutrino mass matrix up to order $\mathcal{O}(\varepsilon ^{2})$.} $\mathcal{%
	O}(\varepsilon ^{2})$%
\begin{equation}
	\begin{array}{c}
		\left\vert m_{1}\right\vert =\sqrt{\left( a+b\right) ^{2}-\left\vert
			\varepsilon \right\vert \left( \cos \phi _{\varepsilon }\right) \left(
			a-b\right) }~,~\left\vert m_{2}\right\vert =\sqrt{\left( a+b\right)
			^{2}+2\left\vert \varepsilon \right\vert \cos \phi _{\varepsilon }\left(
			a+b\right) } \\
		\left\vert m_{3}\right\vert =\sqrt{\left( a-b\right) ^{2}-\left\vert
			\varepsilon \right\vert \left( \cos \phi _{\varepsilon }\right) \left(
			a-b\right) }%
	\end{array}
	\label{ev}
\end{equation}%
where the following conditions for the diagonalization must be fulfilled
\begin{equation}
	\tan 2\theta =\sqrt{3}\frac{\left\vert \varepsilon \right\vert \sqrt{%
			a^{2}\sin ^{2}\phi _{\varepsilon }+b^{2}\cos ^{2}\phi _{\varepsilon }}}{%
		\left\vert \varepsilon \right\vert b\cos \phi _{\varepsilon }-2ab}\quad
	,\quad \tan \sigma =\frac{a}{b}\tan \phi _{\varepsilon }\
\end{equation}%
From these masses, we deduce the expressions for the solar and atmospheric
mass-squared differences \
\begin{equation}
	\Delta m_{21}^{2}=\left\vert \varepsilon \right\vert \left( 3a+b\right) \cos
	\phi _{\varepsilon }\quad ,\quad \Delta m_{31}^{2}=-4ab  \label{sa}
\end{equation}%
As for the mixing angles $\theta _{13}$, $\theta _{23}$ and $\theta _{12}$,
they may be expressed in the case of trimaximal mixing as a function of $%
\theta $ and $\sigma $, we have%
\begin{equation}
	\sin ^{2}\theta _{13}=\frac{2}{3}\sin ^{2}\theta \quad ,\quad \sin
	^{2}\theta _{12}=\frac{1}{3-2\sin ^{2}\theta }\quad ,\quad \sin ^{2}\theta
	_{23}=\frac{1}{2}-\frac{\sqrt{3}\sin 2\theta }{2\left( 3-\sin ^{2}\theta
		\right) }\cos \sigma  \label{ma}
\end{equation}%
From the $3\sigma $ allowed range of $\sin ^{2}\theta _{13}$ \cite{R6} and
the first equation in (\ref{ma}), we find that the range of $\theta $ is
given by $0.176\lesssim \theta \left[ \mathrm{rad}\right] \lesssim 0.193$.
In the left panel of Fig. \ref{01}, we show the correlation among the parameters $a$, $b$ and $\varepsilon $ where the displayed points satisfy the the $3\sigma $\ experimental values of the oscillation parameters. It is clear from this figure that $a$ and $b$ have opposite signs\textrm{\ }which means that---according to eq. (\ref{sa})---$\Delta m_{31}^{2}>0$ and thus, the current model predicts the normal hierarchy for neutrino masses which will be focused on in our numerical study. In the right panel of Fig. \ref{01}, we display the atmospheric angle as a function of $\varepsilon $ with the palette showing $\phi _{\varepsilon }$. The obtained ranges of these
parameters are%
\begin{eqnarray}
	0.41503 &\lesssim &\sin ^{2}\theta _{23}\lesssim 0.49676\quad ,\quad
	0.95779\lesssim \phi _{\varepsilon }\left[ \mathrm{rad}\right] \lesssim
	2.18245  \nonumber \\
	\varepsilon &\in &\left[ -0.01363\rightarrow -0.00597\right] \cup \left[
	0.00597\rightarrow 0.01363\right]
\end{eqnarray}%
It is clear from this figure that the atmospheric angle lies in the lower
octant ($\theta _{23}<\pi /4$) which is one more important prediction of the
model. Another way of getting this prediction is by using the equation of
the atmospheric angle in (\ref{ma}).
In this regard, by using the interval of $\theta $ as an input and we allow
the arbitrary phase $\sigma $ to vary randomly in the interval $[-\pi ,\pi ]$
along with the expressions of the mixing angles in eq. (\ref{ma}) and the $%
3\sigma $ allowed ranges of neutrino oscillation parameters from Ref. \cite%
{R6}, we show\ in the left panel of Fig. \ref{02} the correlation between $%
\sin ^{2}\theta _{23}$ and $\cos \sigma $\ from which we extract the
constrained range of $\sigma $ given as
\begin{equation}
	\sigma \in \left[ -1.5406\rightarrow 1.5415\right]
\end{equation}%
It is clear from this plot that $\cos \sigma >0$ and thus, by\ plugging the
values of $\theta $ and $\cos \sigma $\ back in the expression of the
atmospheric angle, we get the desired result; $\sin ^{2}\theta _{23}<1/2$.
On the other hand, a relationship between the arbitrary phase $\sigma $\ and
the Dirac phase $\delta _{CP}$\ can be obtained by means of the Jarlskog
invariant parameter defined as $J_{CP}=Im(U_(e1)U_(\mu1)^(*)U_(\mu2)U_(e2)^(*))$. In the PDG parametrization\emph{\ }\cite%
{R9},\emph{\ }$J_{CP}$\emph{\ }is given in terms of the mixing angles\ and
the Dirac $CP$ phase as
\begin{equation}
	J_{CP}=\frac{1}{8}\sin 2\theta _{12}\sin 2\theta _{13}\sin 2\theta _{23}\cos
	\theta _{13}\sin \delta _{CP},  \label{jc1}
\end{equation}
while in the case of trimaximal mixing matrix (\ref{tm2}), $J_{CP}$ takes
the following form\emph{\ }%
\begin{equation}
	\left. J_{CP}\right\vert _{TM_{2}}=\left( 1/6\sqrt{3}\right) \sin 2\theta
	\sin \sigma  \label{jc2}
\end{equation}%
In the right panel of Fig. \ref{02}, we plot the Jarlskog invariant
parameter as a function of $\varepsilon $\ with the color code indicating
the range of the Dirac $CP$ phase $\delta _{CP}$\ where we find that $J_{CP}$%
\ is predicted to be scattered in the region $-0.0357\lesssim J_{CP}\lesssim
0.0302$.\emph{\ }Furthermore, by using eq. (\ref{ma}) and identifying the%
\emph{\ }Jarlskog parameter from the PDG parametrization with the one from
trimaximal mixing, we obtain\emph{\ }a relation between\emph{\ }$\sigma $
and $\delta _{CP}$
\begin{equation}
	\sin \delta _{CP}=\sin \sigma /\sin 2\theta _{23}
\end{equation}%
From this equation, we deduce that $\sigma $ and $\delta _{CP}$\ are always
different from the $CP$ conserving values $n\pi $ where $n$ can be any
integer and therefore, it is easy to check from eqs. (\ref{jc1}) and (\ref%
{jc2}) that the Jarlskog invariant parameter does not vanish and
consequently, $CP$ violation always occurs in the present model.

\section{Phenomenological consequences}

Besides neutrino oscillation experiments, information on the absolute
neutrino mass scale can be obtained using three different sources:

\begin{itemize}
	\item $\emph{(1)}$ Constraints from cosmological observations providing an
	upper bound on the sum of the three active neutrino masses; $\sum
	m_{i}=m_{1}+m_{2}+m_{3}$. The present upper bound on $\sum m_{i}$ from the
	Planck collaboration is given by $\sum m_{i}<0.12$ $\mathrm{eV}$ at 95\% C.L
	\cite{R14}.
	
	\item \emph{(2)} Direct determination of the neutrino mass by measuring the
	energy spectrum of electrons produced in the $\beta $-decay of nuclei which
	allows to get information on the effective electron antineutrino mass
	defined by
	\begin{equation}
		m_{\beta }=\left( \sum_{i=1,2,3}m_{i}^{2}\left\vert U_{ei}\right\vert
		^{2}\right) ^{1/2}.  \label{ne}
	\end{equation}%
	The current limit from tritium beta decay is given by the KATRIN project,
	which aims at a detection of $m_{\beta }$ with a sensitivity of $0.2$ $%
	\mathrm{eV}$ \cite{R15}.
	
	\item $\emph{(3)}$ Search for $0\nu \beta \beta $ decay processes having a
	decay amplitude proportional to the effective Majorana neutrino mass defined
	as%
	\begin{equation}
		\left\vert m_{\beta \beta }\right\vert =\left\vert
		\sum_{i=1,2,3}U_{ei}^{2}m_{i}\right\vert .  \label{mee}
	\end{equation}%
	This is also considered as the unique probe for the Majorana nature of
	neutrinos. There are many ongoing and upcoming experiments such as GERDA
	\cite{R16}, CUORE \cite{R17}, KamLand-Zen \cite{R18}, GERDA Phase II \cite{R19}, nEXO
	\cite{R20}, which aim to achieve a sensitivity up to $0.01$ $\mathrm{eV}$
	for $\left\vert m_{\beta \beta }\right\vert $.
	\begin{figure}
		\centering
		\hspace{2.5em} \includegraphics[width=.44\textwidth]{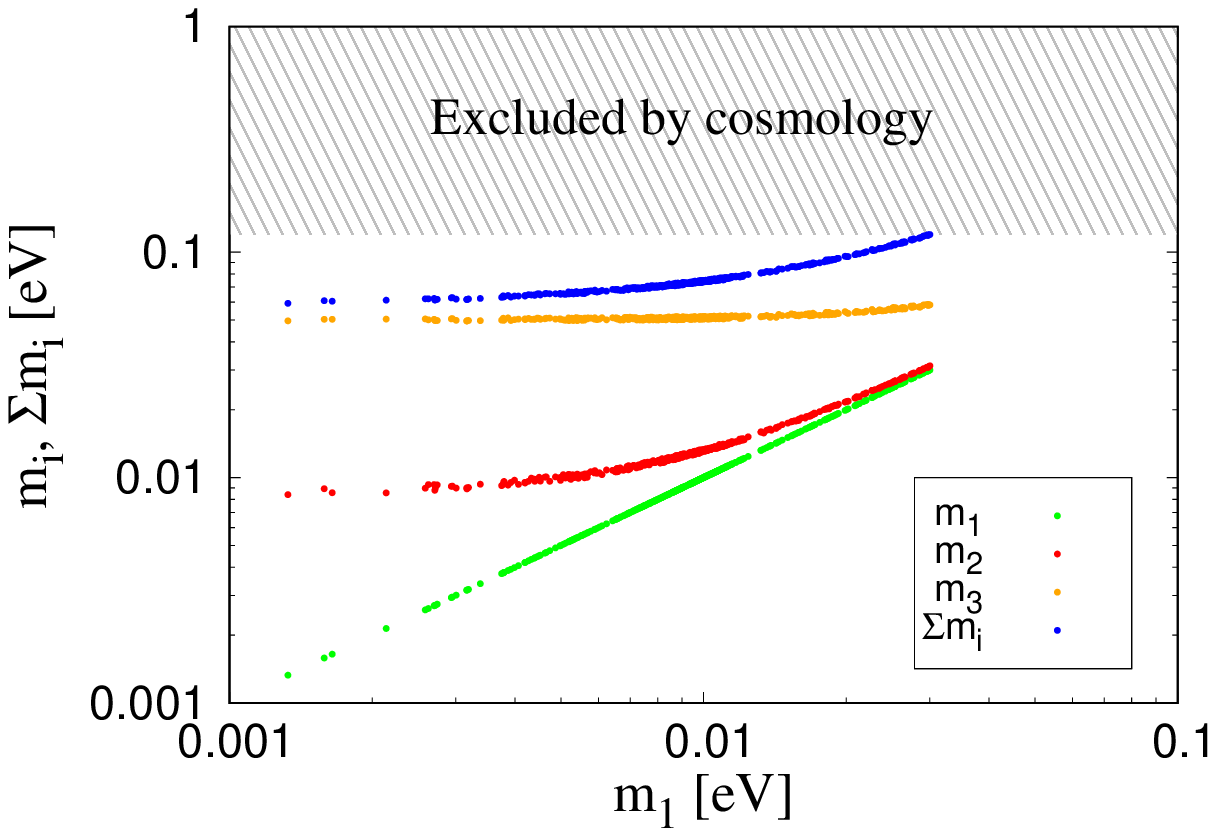}\quad %
		\includegraphics[width=.44\textwidth]{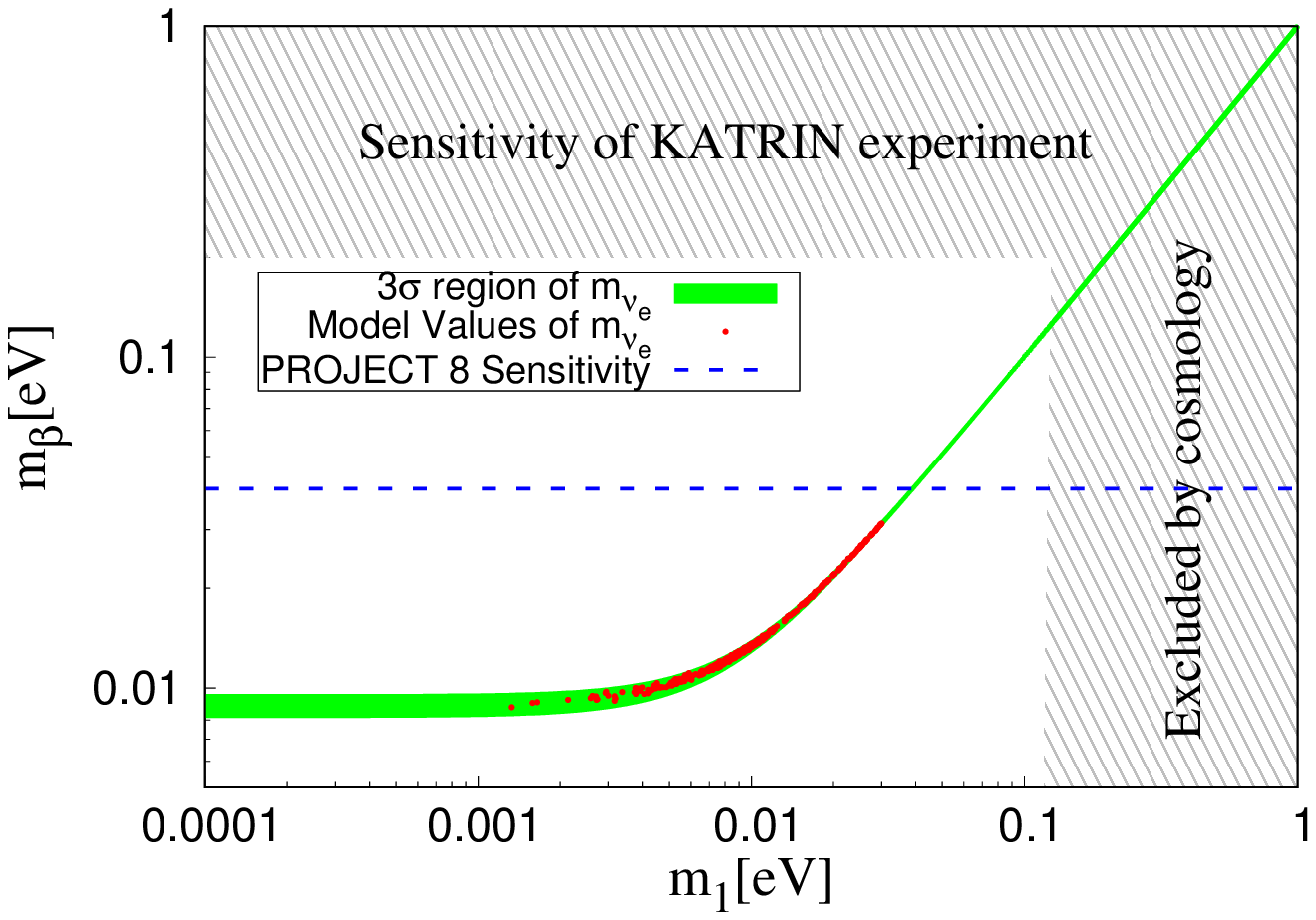}
		\par
		\caption{Left: Predictions for the absolute neutrino masses $m_{i=1,2,3}$
			and their sum $\sum m_{i}$\ as a function of the lightest neutrino mass $%
			m_{1}$. The horizontal filled area represent the upper limit on $\sum m_{i}$
			from Planck collaboration. Right: $m_{\protect\beta}$ as a function of $%
			m_{1} $. The vertical (horizontal) filled area is disfavored by Planck+BAO
			(KATRIN ) data.}
		\label{03}
	\end{figure}
\end{itemize}

In the NH case where $m_{1}\lesssim m_{2}<m_{3}$, it is useful to rewrite
the heaviest states in terms of the lightest one and the mass squared
differences as $m_{2}=\sqrt{m_{1}^{2}+\Delta m_{21}^{2}}$ and\ $m_{3}=\sqrt{%
	m_{1}^{2}+\Delta m_{31}^{2}}$, where $m_{1}$, $m_{2}$ and$\ m_{3}$ are as
given in eq. (\ref{ev}). By using the $3\sigma $ allowed ranges of neutrino
oscillation parameters from Ref. \cite{R6},\textrm{\ }we show in the left
panel of Fig. \ref{03} the correlation between\textrm{\ }the sum $\sum m_{i}$
and the three neutrino masses $m_{i=1,2,3}$ as a function of the lightest
neutrino mass\textrm{\ }$m_{1}$. The predicted regions are as follows%
\begin{eqnarray}
	0.00132 &\lesssim &m_{1}\left[ \mathrm{eV}\right] \lesssim 0.03002\ \ ,\ \
	0.00840\lesssim m_{2}\left[ \mathrm{eV}\right] \lesssim 0.03130  \nonumber \\
	0.04948 &\lesssim &m_{3}\left[ \mathrm{eV}\right] \lesssim 0.05871\ \ ,\ \
	0.05928\lesssim \sum m_{i}\left[ \mathrm{eV}\right] \lesssim 0.11966
	\label{su}
\end{eqnarray}%
We find that the obtained range of $\sum m_{i}$\ is below the upper bound
given by the Planck collaboration with a lower bound given by\textrm{\ }$%
\sum m_{i}\gtrsim 0.05928$ $\mathrm{eV}$. Even though this lower limit is
small, it can be tested by future experiments such as CORE+BAO aiming for a
bound on $\sum m_{i}$ around $0.062\mathrm{eV}$ \cite{R21}. In the right
panel of Fig. \ref{03}, we show the allowed region of the lightest neutrino
mass $m_{1}$ and the effective electron antineutrino mass $m_{\beta }$. In
this plot, we first used the definition of $m_{\beta }$\ given in eq. (\ref%
{ne}) and we replaced $U_{ei}$ by the elements of the first row of TM$_{2}$
given in eq. (\ref{tm2}). Then, by taking into consideration the $3\sigma $
allowed ranges of neutrino oscillation parameters as well as the neutrino
mass constraints in eq. (\ref{su}), we obtain the following range for $%
m_{\beta }$
\begin{equation}
	0.00873\lesssim m_{\beta }\left[ \mathrm{eV}\right] \lesssim 0.03135
	\label{mb}
\end{equation}%
The values in this range are too small compared to the KATRIN sensitivity ($%
\sim 0.2$ $\mathrm{eV}$) and the expected sensitivity from HOLMES experiment
($\sim 0.1$ $\mathrm{eV}$) \cite{R22}, and thus, they can not be tested in
the foreseeable future. On the other hand, as shown by the blue dashed line
in the right panel of Fig. \ref{03}, the upper limit in \ref{mb} is close to
the anticipated sensitivity of Project 8 collaboration ($0.04$ $\mathrm{eV}$%
) \cite{R23} and hence, can be reached by future experiments targeting a
more enhanced sensitivities. \newline
\begin{figure}
	\centering
	\includegraphics[width=.5\textwidth]{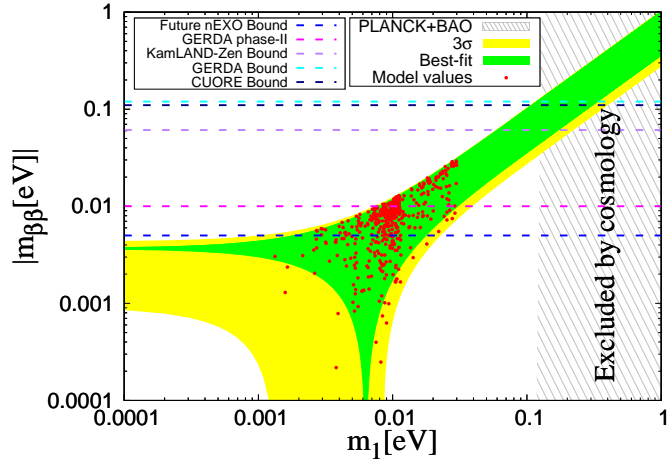}\quad
	\par
	\caption{$\left\vert m_{\protect\beta \protect\beta }\right\vert $ as a
		function of $m_{1}$. The vertical filled area presents the upper limit on
		the sum of the three light neutrino masses from Planck+BAO data.}
	\label{04}
\end{figure}
By repeating the same exercise as before, we first use the definition of $%
\left\vert m_{\beta \beta }\right\vert $\ given in eq. (\ref{mee}) and we
replaced $U_{ei}$ by the elements of the first row of TM$_{2}$ given in eq. (%
\ref{tm2}). Then, we plot in Fig. \ref{04} the the effective Majorana
neutrino mass $\left\vert m_{\beta \beta }\right\vert $ as a function of the
lightest neutrino mass $m_{1}$ where the Majorana phases $\alpha _{21}$ and $%
\alpha _{31}$ are allowed to vary in the range $\left[ 0\rightarrow 2\pi %
\right] $. The horizontal dashed lines in this figure denote the $\left\vert
m_{\beta \beta }\right\vert $ sensitivity from several $0\nu \beta \beta $
decay experiments. Our predicted values (red dots) for $\left\vert m_{\beta
	\beta }\right\vert $ that satisfy constraints from oscillation experiments
are given by the following range%
\begin{equation}
	0.00021\lesssim \left\vert m_{\beta \beta }[\mathrm{eV}]\right\vert \lesssim
	0.02923
\end{equation}%
From this figure, we observe that all the allowed points are below the
current sensitivities from GERDA, CUORE, and KamLand-Zen experiments, while
most of these points are scattered within the future upper bounds of GERDA Phase II and
nEXO experiments, and thus, our predicted values of $\left\vert m_{\beta
	\beta }\right\vert $ can be tested by these experiments.

\section{Conclusion}

In this work, we have constructed a neutrino flavor model based on the $%
\Delta \left( 54\right) $ discrete group that accommodates the observed
neutrino masses and mixing. We have shown that the combination of type II
seesaw mechanism with $\Delta \left( 54\right) $ flavor group in the same
framework lead to a simple neutrino mass matrix with one texture zero which
restricts the number of free parameters to three parameters. Moreover, the
obtained neutrino mass matrix has the magic symmetry and thus, predicts a
rich phenomenology provided by the trimaximal mixing of neutrinos. In
particular, the most interesting results of our model is that it predicts
the normal hierarchy for neutrino masses and the lower octant for the
atmospheric angle. We have studied the phenomenological implications
associated with neutrino masses where we found that the range of the
effective Majorana neutrino mass $m_{\beta \beta }$ is within the reach of
future experiments while the obtained range of the electron neutrino mass%
\textrm{\ }$m_{\beta }$ is far from current and future experimental bounds.
Many issues remain to be studied in this promising model, such as the quark
and the scalar sectors and their phenomenology. On the other hand, since we
found that $CP$ is always violated in the current model, it would be useful
to examine other neutrino related topics like the triplet contribution to
the lepton asymmetry of the Universe; we leave detailed investigations of
these subjects to future work.

\end{document}